\documentclass[aps,prb,twocolumn,superscriptaddress,showpacs]{revtex4}

\usepackage{color}
\usepackage{graphicx}

\begin{document}
\input epsf.sty
\title{Soft phonons and structural phase transitions in 
La$_{1.875}$Ba$_{0.125}$CuO$_{4}$}


\author{H.~Kimura}
\email[]{kimura@tagen.tohoku.ac.jp}
\affiliation{Institute of Multidisciplinary Research for Advanced Materials, 
Tohoku University, Sendai 980-8577, Japan}


\author{Y.~Noda}
\affiliation{Institute of Multidisciplinary Research for Advanced Materials, 
Tohoku University, Sendai 980-8577, Japan}

\author{H.~Goka}
\affiliation{Institute for Chemical Research, Kyoto University, Gokasho, Uji 610-0011, Japan}

\author{M.~Fujita}
\affiliation{Institute for Material Research, Tohoku University, Sendai 980-8577, Japan}

\author{K.~Yamada}
\affiliation{Institute for Material Research, Tohoku University, Sendai 980-8577, Japan}

\author{G.~Shirane}
\affiliation{Department of physics, Brookhaven National Laboratory, Upton, 
New York 11973-5000, U. S. A.}


\date{\today}

\begin{abstract}
Soft phonon behavior associated with a structural phase transition from 
the low-temperature-orthorhombic (LTO) phase ($Bmab$ symmetry) to 
the low-temperature-tetragonal (LTT) phase ($P4_{2}/ncm$ symmetry) was investigated 
in La$_{1.875}$Ba$_{0.125}$CuO$_{4}$ using neutron scattering. As temperature decreases, 
the TO-mode at $Z$-point softens and approaches to zero energy around $T_{\rm d2}=62$~K, 
where the LTO -- LTT transition occurs. Below $T_{\rm d2}$, the phonon hardens quite rapidly 
and it's energy almost saturates below 50~K. At $T_{\rm d2}$, 
the energy dispersion of the soft phonon along in-plane direction significantly changes 
while the dispersion along out-of-plane direction is 
almost temperature independent. 
Coexistence between the LTO phase and the LTT phase, 
seen in both the soft phonon spectra and the peak profiles of Bragg reflection, is discussed 
in context of the order of structural phase transitions. 
\end{abstract}


\maketitle

\section{Introduction}\label{intro}
\begin{figure}[bp]
\centerline{\epsfxsize=3in\epsfbox{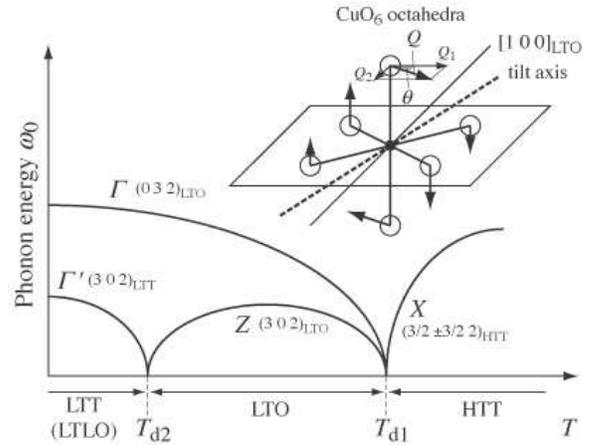}}
\caption{Schematic diagram for the energy of soft phonons as a function of temperature. 
Inset depicts the definitions for the tilting of CuO$_{6}$ octahedra.}
\label{fig1}
\end{figure}

It is well known that La-214 based high-$T_{\rm c}$ cuprates 
undergo successive structural phase transitions. On cooling temperature, 
the second-order phase transition from the high-temperature tetragonal 
(HTT; $I4/mmm$) to the low-temperature orthorhombic (LTO; $Bmab$) occurs. 
In some cases, as seen in the 
La$_{2-x}$Ba$_{x}$CuO$_{4}$ (LBCO)\cite{Moodenbaugh1988,Axe1989} or the 
La$_{2-x-y}$Nd$_{y}$Sr$_{x}$CuO$_{4}$ (LNSCO)\cite{Crawford1991} 
systems, further structural phase transitions 
from the LTO phase to the low-temperature-less orthorhombic (LTLO; $Pccn$) or to the 
low-temperature tetragonal (LTT; $P4_{2}/ncm$) follow the HTT-LTO transition 
with decreasing temperature. 
These structural phases can be categorized by the pattern of coherent tilting for 
CuO$_{6}$ octahedra. 
Extensive neutron scattering studies on La$_{2-x}$Sr$_{x}$CuO$_{4}$ (LSCO) 
have revealed that a condensation of zone-boundary soft phonons causes 
the HTT-LTO structural phase transition, resulting in 
a staggered tilting of CuO$_{6}$ octahedra\cite{Birgeneau1987,Boni1988,Thurston1989}. 
At the HTT-LTO transition, a degenerate pair of 
TO-phonon at $X$-point wave vectors ${\bf q}=(1/2\ \pm1/2\ 0)_{\rm HTT}$, 
associated with the tilting motion of CuO$_{6}$ octahedra, 
goes soft and split into two modes at ${\it \Gamma}$- and $Z$-points in 
the LTO phase. In LSCO system, $Z$-point phonon once hardens just below the transition 
temperature but softens again with decreasing 
temperature, indicating the structural instability 
toward the LTO-LTLO or LTO-LTT transition\cite{Thurston1989}. 
Keimer {\it et al}. have experimentally shown in La$_{1.65}$Nd$_{0.35}$CuO$_{4}$ that 
the LTO-LTLO transition is also understood as 
the ordinary displacive phase transition with the softening of 
$Z$-point phonons, which turns into zone-center phonon in the LTLO phase\cite{Keimer1993} 
(We call this phonon as ${\it \Gamma}^{\prime}$-point for the sake of convenience). 
The summary of the soft phonon behaviors 
at several structural phases and definitions of order parameter involving 
\{$Q_{1}, Q_{2}$\} or \{$Q, \theta$\} for the tilting of CuO$_{6}$ 
octahedra\cite{Kimura2004} are schematically shown in Fig.~\ref{fig1}. 

The relevance between the structural instability and 
the high-$T_{\rm c}$ superconductivity has been also studied intensively in these systems. 
It has been reported\cite{Braden1993} that the evolution of 
the order parameter of the LTO phase 
suppressed below $T_{\rm c}$ in the LSCO of $x=0.13$. 
After that it has been discovered\cite{Lee1996,Kimura2000} that 
the softening of $Z$-point phonons {\em breaks} at $T_{\rm c}$ in 
the LSCO of $x=0.15$ and 0.18. 
There facts indicate a competition between the LTO-LTLO (or LTO-LTT) 
structural instability and the high-$T_{\rm c}$ superconductivity. 
This competitive relation becomes 
more apparent in the LBCO and LNSCO systems around 1/8-doping, where the charge stripe 
order follows on the LTO-LTT phase transition, 
resulting in the suppression of superconductivity\cite{Tranquada1995,Tranquada1996}. 
The motivation of the phonon studies in LBCO system is to clarify 
the relevance among the low-energy lattice dynamics in the LTT phase, 
the high-$T_{\rm c}$ superconductivity, and the charge stripe order. 

In the present study, we examined the soft phonon behavior in 1/8-hole doped LBCO 
for a wide temperature range including the HTT-LTO and the LTO-LTT structural 
phase transition. Our measurement is a first case which observes 
the soft phonons associated with the first-order LTO-LTT transition in La-214 system. 
\section{Experimental details}\label{exp}
\begin{figure}[bp]
\centerline{\epsfxsize=3in\epsfbox{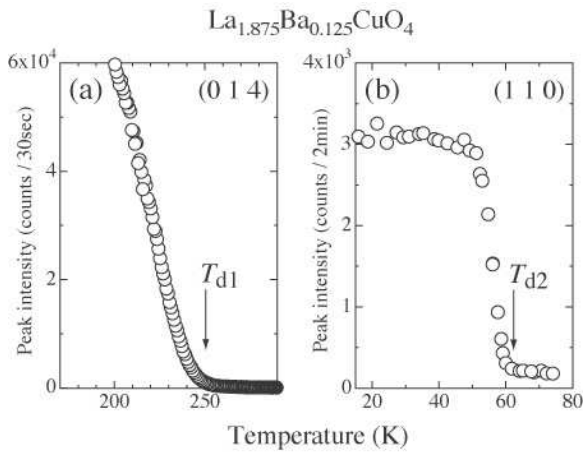}}
\caption{Temperature dependence of (a); $(0\ 1\ 4)$- and (b); 
$(1\ 1\ 0)$-superlattice intensities, which corresponds to the order parameter 
for the LTO and the LTT phase, respectively.}
\label{fig2}
\end{figure}

The single crystal of La$_{1.875}$Ba$_{0.125}$CuO$_{4}$ was grown by 
traveling-solvent floating-zone method. Details in growing the single crystal 
are described elsewhere\cite{Fujita2004}. 
The crystal has a cylindrical shape of which dimension is $\sim8$~mm in diameter with 
20~mm height. Note that the crystal in the present study is the very same as 
used in the measurements of spin dynamics related with the charge stripe order\cite{Fujita2004}.
Diamagnetic susceptibility was measured with SQUID magnetometer, showing no 
bulk superconductivity down to $T=2$~K. We determined the structural phase transition 
temperatures for the HTT-LTO (defined as $T_{\rm d1}$) and for 
the LTO-LTT (defined as $T_{\rm d2}$) transitions, which is quite 
sensitive to the Ba concentration in the crystal. 
Figures~\ref{fig2}(a) and (b) are the temperature variations 
of peak intensities for $(0\ 1\ 4)$ and $(1\ 1\ 0)$ Bragg reflections 
measured with neutron diffraction, which 
corresponds to the order parameters for the LTO and the LTT phase, respectively. 
As a result, $T_{\rm d1}$ and $T_{\rm d2}$ were obtained as 250~K and 62~K, respectively, 
which is almost consistent with the previous results\cite{Axe1989}. 
Note that all the Bragg reflections measured at the lowest temperature ($\sim15$~K) 
had well defined single peak profiles, showing that 
the crystal structure in the lowest temperature phase is not a orthorhombic but 
a tetragonal. The lattice parameters obtained at $T=16$~K are $a=b=5.351$~\AA\ 
and $c=13.239$~\AA. 

Inelastic neutron scattering measurements of soft phonons 
were performed at triple-axis spectrometer TOPAN of Tohoku University installed at 
JRR-3M reactor in Japan Atomic Energy Research Institute (JAERI). 
The final energy of neutrons was fixed at $E_{\rm f}=13.5$~meV using 
PG(002) analyzer. The sequences of the horizontal collimations were 
40'-30'-30'-60' and 30'-30'-30'-60', which yields the energy resolutions of 
0.9~meV and 0.8~meV, respectively. A pyrolytic graphite filter and a sapphire cut-off filter 
were inserted for scattered and incident beam to reduce the higher-order contamination 
of neutrons. The sample was mounted in the $(h\ 0\ l)$ zone. 
Since the single crystal has twinned domains at the LTO phase, 
both the $(h\ 0\ l)$ and $(0\ k\ l)$ zones are superposed. Throughout this paper, 
if no specification is given, the reciprocal space is described 
by the reciprocal lattice unit in the LTO ($Bmab$) coordination, 
of which definitions are consistent with that in the LTT ($P4_{2}/ncm$) one. 
Temperature of the sample is controlled between 16~K and 450~K using 
$^{4}$He-closed-cycle cryostat with high-power heater. 
\section{Results and Discussion}\label{res}
\subsection{Soft phonon behavior in wide temperature range}\label{res1}
We measured the energy spectra of phonons around $Q=(3\ 0\ 2)$, 
which corresponds to TO-mode at $Z$-point in the LTO phase, 
as a function of temperature and as a function of $Q$. All the scans were performed 
in a constant-$Q$ mode. Since soft phonons become overdamped in 
the vicinity of the structural phase transition 
temperature, we could not obtain a well-defined phonon spectra just at 
$Q=(3\ 0\ 2)$ (=$Z$-point). Therefore we measured 
$(3+q\ 0\ 2)$ mode and determined the phonon dispersion as a function of temperature. 

\begin{figure}[tp]
\centerline{\epsfxsize=3in\epsfbox{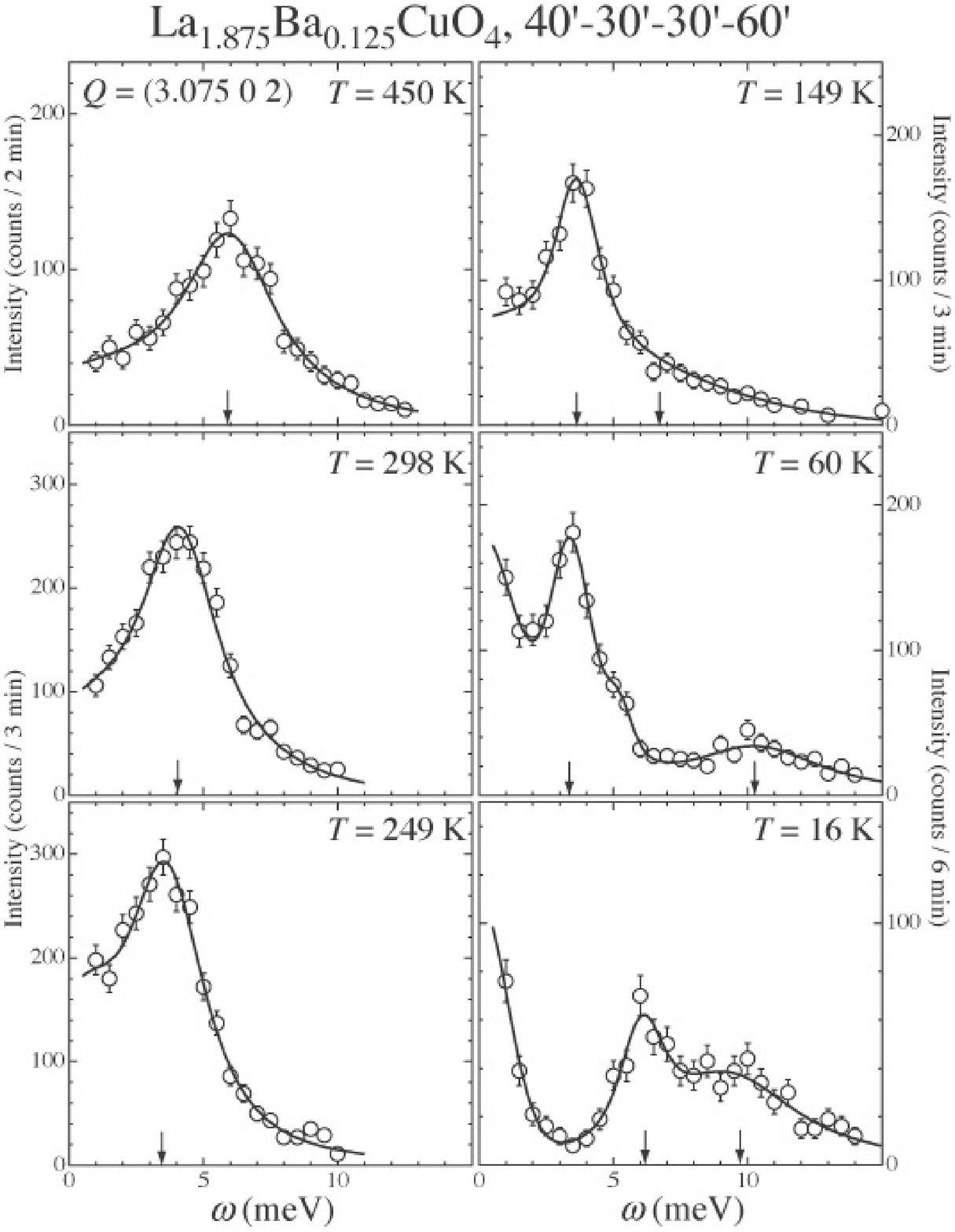}}
\caption{Energy spectra of $(3.075\ 0\ 2)$ phonons at several temperatures.}
\label{fig3}
\vfill\vspace*{17mm}
\centerline{\epsfxsize=2.3in\epsfbox{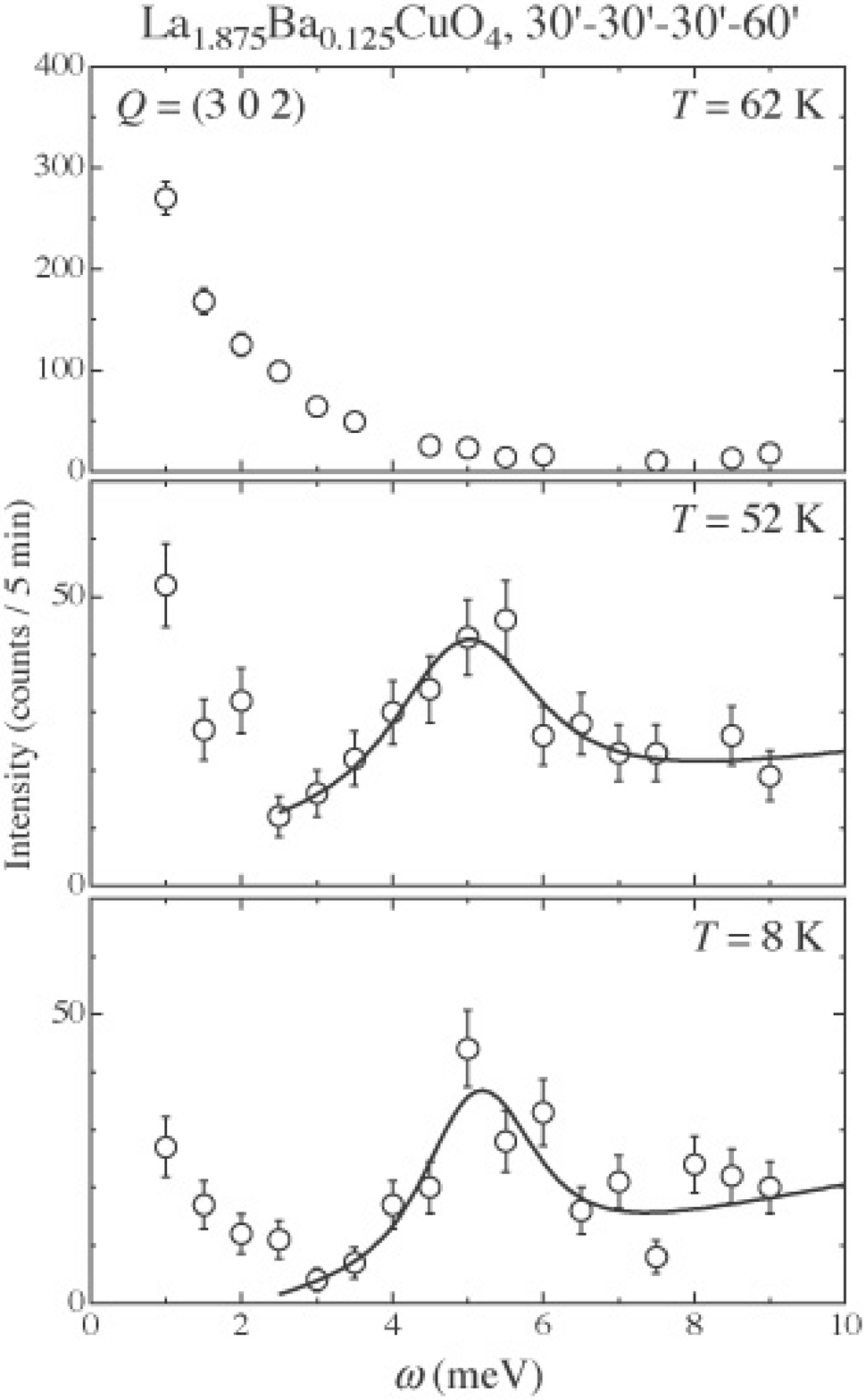}}
\caption{Energy of the four phonon modes as a function of temperature. Lines in the figure 
are guides to the eye.}
\label{fig4}
\end{figure}

Typical energy spectra of the phonons at $Q=(3.075\ 0\ 2)$ 
are shown in Fig~\ref{fig3}, taken at several temperatures. 
The phonon energy $\omega_{\rm ph}$ was extracted by fitting the observed spectra 
with the functions of a dumped harmonic oscillator form and 
a Gaussian central peak, which convolved with the instrumental resolution. 
In these fittings, we assumed $\omega_{\rm ph}$ to be constant within 
the energy resolution, meaning that the dispersion of phonons was not considered. 
The solid lines in Fig~\ref{fig3} are the results of fittings, which well reproduce 
the observed spectra. The arrows in the figures point to 
the peak positions determined by the fittings. 
Above $T_{\rm d1}$ ($=250$~K), a single phonon peak 
is clearly seen, which corresponds to the phonon branch from the degenerate 
$X$-point in the HTT phase. 
$\omega_{\rm ph}$ decreases as temperature decreases, indicates that 
the $X$-point phonon goes soft toward the HTT-LTO phase transtion. 
Below $T_{\rm d1}$, in the LTO phase, two phonon peaks appear, 
showing that the degenerate $X$-point phonon split into 
a lower energy phonon at $Z$-point ($=(3\ 0\ 2)$) and a higher one 
at ${\it \Gamma}$-point ($=(0\ 3\ 2)$). Note that the two phonons are observable 
simultaneously because of the twinning of $(h\ 0\ l)$ and $(0\ k\ l)$ zones 
in the LTO phase. 
As seen in the figures, $\omega_{\rm ph}$ for $Z$-point at $T=60$~K increases at 
$T=16$~K. This indicates that the LTO-LTT structural phase transitions 
occurs and the $Z$-point phonons turn into 
${\it \Gamma}^{\prime}$-point around $T=60$~K, 
which is consistent with the temperature dependence of 
$(1\ 1\ 0)$ super lattice shown in Fig.~\ref{fig2} (b). 
Figure~\ref{fig4} shows the representative phonon spectra just at $Q=(3\ 0\ 2)$, 
taken below $T_{\rm d2}$. At $T=62$~K, we cannot see clear phonon peak because of the 
overdamping of phonons in the vicinity of the structural phase transition. However at $T=52$~K, 
below $T_{\rm d2}$, the phonon peak becomes well defined around $\sim5$~meV 
and its phonon energy is almost temperature independent down to 16~K. 

\begin{figure}[b]
\centerline{\epsfxsize=3in\epsfbox{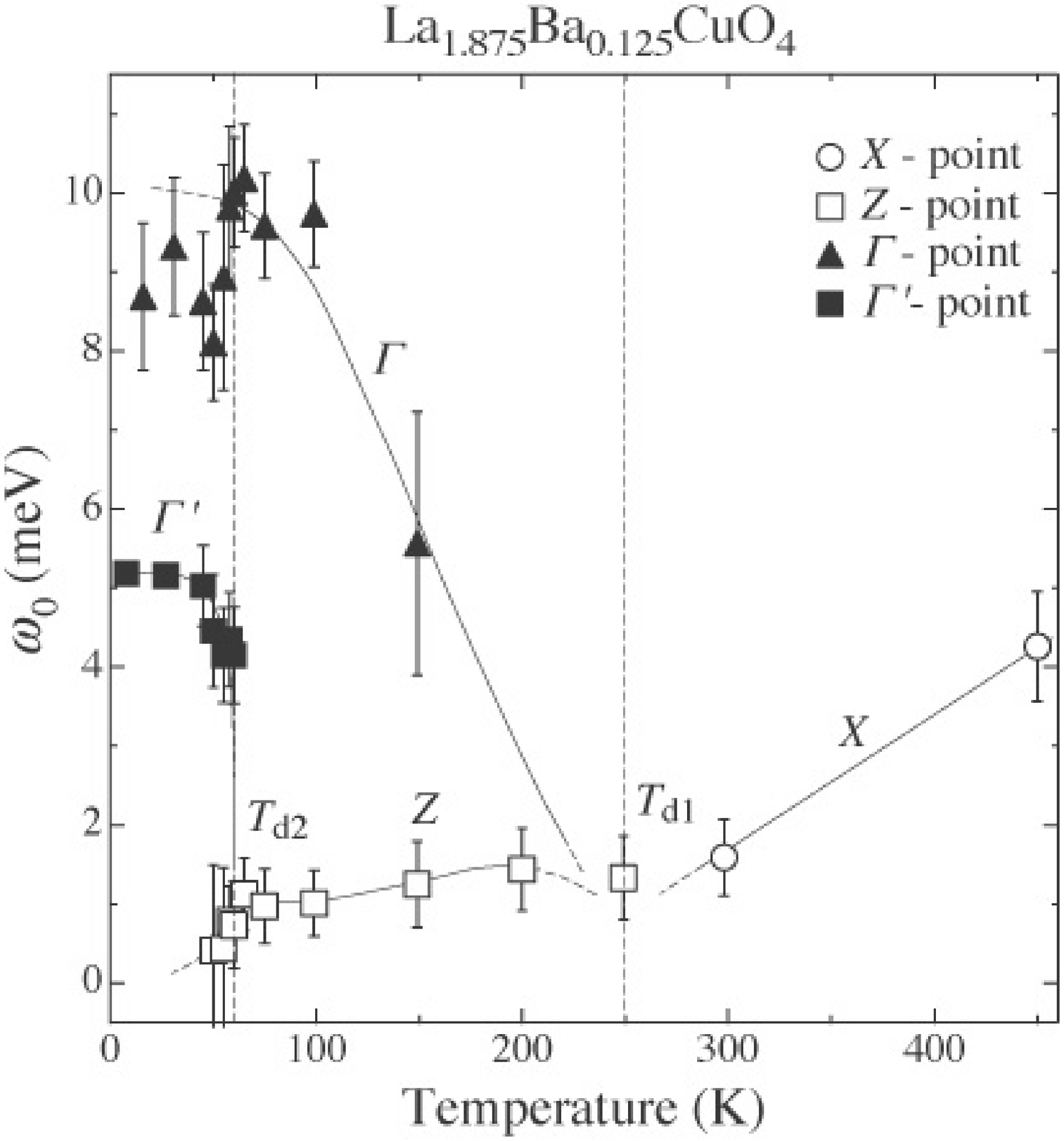}}
\caption{Energy of the four phonon modes as a function of temperature. Lines in the figure 
are guides to the eye.}
\label{fig5}
\end{figure}

Figure~\ref{fig5} shows the summary of soft phonon behaviors as 
a function of temperature. The phonon energy at $q=0$, defined as $\omega_{0}$, 
was extrapolated from the dispersion relations of $(3+q\ 0\ 2)$ phonons. 
Note that $\omega_{0}$ for $(3\ 0\ 2)$ ${\it \Gamma}^{\prime}$-point below 60~K 
(denoted by filled squares in Fig~\ref{fig4}) was directly derived from the 
$(3\ 0\ 2)$ phonon spectra. 
The phonon at degenerate $X$-point (open circles) in the HTT phase 
goes soft with decreasing temperature and splits into two modes below 
$T_{\rm d1}$, which is due to 
the reduction of crystal symmetry into orthorhombic with $Bmab$ space group. Thus 
these two modes correspond to zone-boundary phonon at $(3\ 0\ 2)$ $Z$-point 
and zone-center one at $(0\ 3\ 2)$ ${\it \Gamma}$-point. 
On further cooling temperature, the $Z$-point phonon (open squares) 
remains soft and start decreases again in energy while 
the ${\it \Gamma}$-point phonon (filled triangles) keeps on 
hardening with decreasing temperature, indicating that 
the LTO phase evolves and the lattice instability toward the further structural phase 
transition increases. These behaviors are qualitatively consistent 
with that of La$_{2}$CuO$_{4}$\cite{Thurston1989}. 
Below $T\sim60$~K, $Z$-point phonon discontinuously hardens 
and its energy becomes almost temperature independent (filled squares), 
showing that the LTO-LTT phase transition occurs and the LTT phase 
becomes robust rapidly. The phonons at $(0\ 3\ 2)$ 
${\it \Gamma}$-point (denoted by filled triangles) 
show small anomaly around the transition temperature $T_{\rm d2}$, 
which indicates that the fluctuation for the tilt amplitude of CuO$_{6}$ octahedra 
was also affected by the LTO-LTT transition. 
Below $T_{\rm d2}$, $Z$-point becomes zone-center ${\it \Gamma}^{\prime}$-point 
and Raman active, because the structural transition from $B$-base centered lattice to 
primitive lattice occurs. It should be noted that 
the overall behaviors of soft phonons in the present LBCO is 
qualitatively consistent with the results for La$_{1.65}$Nd$_{0.35}$CuO$_{4}$, 
which shows the second order HTT--LTO--LTLO phase transitions\cite{Keimer1993}. 

We systematically measured the phonon spectra of $(3+q\ 0\ 2)$ and $(3\ 0\ 2-q)$ modes 
as a function of $q$ in whole temperature range below 450~K 
to determine the dispersion relation of the soft phonons. 
\begin{figure}[bp]
\centerline{\epsfxsize=2.5in\epsfbox{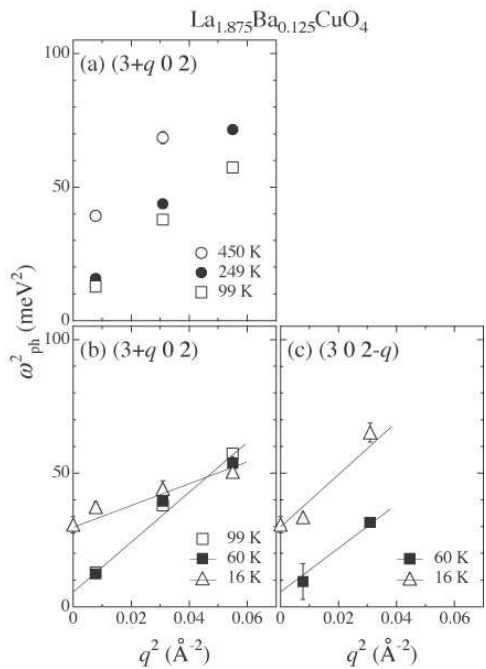}}
\caption{(a), (b); $(3+q\ 0\ 2)$- and (c); $(3\ 0\ 2-q)$-dispersion of the 
soft phonons at several temperatures. Solid lines are guides to the eye.}
\label{fig6}
\end{figure}
We show the dispersion relation as $\omega_{\rm ph}^{2}$ versus $q^{2}$ plot 
because the soft phonon dispersion near a structural phase transition can be expressed as 
\begin{equation}
\omega_{\rm ph}^{2}=A(T-T_{\rm d1(d2)})+b_{1}q_{1}^{2}+b_{2}q_{2}^{2}+b_{3}q_{3}^{2}
\label{eq1}
\end{equation}
where $b_{1}$, $b_{2}$, and $b_{3}$ are coefficients for $a$-, $b$-, and $c$-axis, 
respectively\cite{Thurston1989}. 
Figures~\ref{fig5}(a) and (b) show the dispersion of $(3+q\ 0\ 2)$ mode and 
Fig.\ref{fig5}(c) shows the dispersion of $(3\ 0\ 2-q)$ mode taken at several temperatures. 
Above 60~K, the slope of the dispersion along $a$-axis is 
almost temperature independent and $b_{1}$ was obtained as 1100 meV$^{2}$\AA$^{2}$, 
of which value are roughly consistent with that for 
La$_{1.65}$Nd$_{0.35}$CuO$_{4}$\cite{Keimer1993}. 
Furthermore, as seen in Figs.\ref{fig5}(b) and (c), the dispersion at $T=60$~K along 
$a$-axis is comparable to that along $c$-axis, indicating that the phonons propagate 
isotropically. However, at $T=16$~K, the slope of dispersion along 
$a$-axis significantly changes in comparison with 
that at $T=60$~K, while the the slope of dispersion along $c$-axis almost unchanges 
between 16~K and 60~K. 
This interesting feature 
is characteristic for the LTT structure because the dispersion of the phonon along $a$-axis 
is temperature-independent near the LTO-LTLO structural phase 
transition\cite{Keimer1993} in La$_{1.65}$Nd$_{0.35}$CuO$_{4}$. 
It might be also suggested that the doped carrier makes the phonon velocity 
along $a$-axis reduced and 
there exists a possible relation between this modification 
and the formation of charge stripes in the LBCO system. Further systematic studies 
for soft phonons in this system are required to address this speculation. 
\begin{table}[tbp]
\begin{center}
\setlength{\tabcolsep}{5pt}
\begin{tabular*}{2.1in}{ccc}
 \hline
 $T$ (K)\phantom{0}&$b_{1}\parallel a$ (meV$^{2}$\AA$^{2}$)\phantom{0}&$b_{3}\parallel c$\\ \hline
60&1100&1080\\
16&290&1370\\ \hline
\end{tabular*}
\end{center}
\caption{
The slope of the phonon dispersion parallel to $a$- and $c$-axis at $T=60$~K and 15~K.}
\label{tab1}
\end{table}
The coefficients of the dispersion slope $b_{1}$ and $b_{3}$ 
obtained at $T=60$~K and 16~K are listed in 
Table~\ref{tab1}. 
\subsection{Coexistence between the LTO and the LTT phases}\label{res2}
We focus on the detailed transition scheme from the LTO to the LTT phase. 
Figure~\ref{fig6} is an enlarged figure of Fig.~\ref{fig5} to emphasize the phonon behaviors 
around $T_{\rm d2}$. It is clearly seen that the phonon at $Z$-point (open squares) 
coexists with that at ${\it \Gamma}^{\prime}$-point (filled squares) between 60~K and 50~K, 
which is seen in a region with a gray hatching in the figure. The figure also shows that 
the ${\it \Gamma}^{\prime}$-point phonon discontinuously appears, and the $\omega_{0}$ of 
$Z$-point phonon gets close to zero quite rapidly but remains at finite energy 
below $T_{\rm d2}$, indicating the incomplete softening due to a first-order LTO-LTT 
transition. Note that 
the energy of the ${\it \Gamma}$-point phonon (filled triangle) gradually 
decreases with decreasing temperature where the LTO and the LTT phases coexist. 
\begin{figure}[t]
\centerline{\epsfxsize=3in\epsfbox{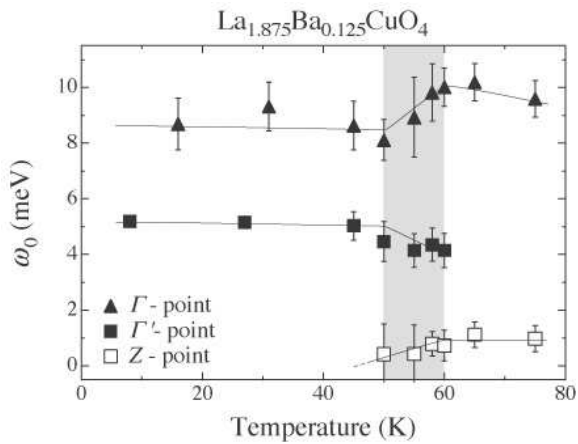}}
\caption{Enlarged view of the energy of the phonon modes as a function of temperature. 
Coexistence between the $Z$-point phonon and the ${\it \Gamma}^{\prime}$-point one is seen in 
the gray hatched region. Several lines are guides to the eye.}
\label{fig7}
\end{figure}

This coexistence is apparently seen not only in the phonon spectra but also 
in the peak profiles of Bragg reflections. 
Figure~\ref{fig7} shows $(3.075\ 0\ 2)$ phonon spectra and $h$-scan profiles 
of $(1\ 1\ 0)_{\rm HTT}$ Bragg reflection taken at three 
different temperatures between 
60~K and 50~K. Mirror index of $(1\ 1\ 0)$ for Bragg reflection is coordinated as 
used in the HTT phase, which corresponds to $(2\ 0\ 0)$ in the LTO 
coordination. Three peaks are seen in the phonon spectra as shown in Figs.~\ref{fig7}(a)-(c). 
The peaks located at the lowest, medium, and the highest energy correspond to 
$(3\ 0\ 2)$ $Z$-, $(3\ 0\ 2)$ ${\it \Gamma}^{\prime}$-, 
and $(0\ 3\ 2)$ ${\it \Gamma}$-points, 
respectively. As temperature decreases, the intensity of the phonon for $Z$-point 
decreases whereas that for ${\it \Gamma}^{\prime}$-point increases, and the 
${\it \Gamma}^{\prime}$-point phonon finally robust at $T=50$~K. 
The present crystal has four domain in the LTO phase, resulting in 
the peak splitting into four $(1\ 1\ 0)_{\rm HTT}$ Bragg points. 
The geometry of the splitting is illustrated in the inset of Fig.~\ref{fig7}(d). 
At 60~K, as shown in Fig.~\ref{fig7}(d), the peak profiles has components of 
two main peaks from the LTO domains and a minor peak at $(1\ 1\ 0)_{\rm HTT}$ 
position which comes from the LTT structure. With decreasing temperature, 
the central peak increases but two sideward peaks decreases, and finally, the peak at 
$(1\ 1\ 0)_{\rm HTT}$ becomes almost dominant at $T=50$~K, of which behaviors has been 
also reported previously\cite{Billinge1993} and 
consistent with the present results for soft phonons. Therefore, we conclude that 
the LTO phase and the LTT phase competitively coexist and 
the volume fraction of each phase changes with decreasing temperature. 

We finally introduce a simple discussion by the framework of 
Landau phenomenological theory\cite{Axe1989,Ting1990} 
\begin{figure}[t]
\centerline{\epsfxsize=2.8in\epsfbox{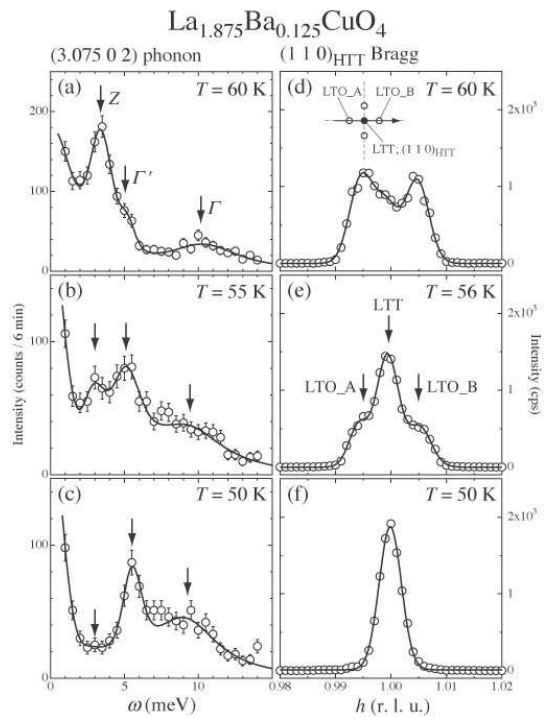}}
\caption{(a)-(c); Energy spectra of $(3.075\ 0\ 2)$ phonons and $h$-scan profiles 
of $(1\ 1\ 0)_{\rm HTT}$ Bragg peak taken at three different temperatures 
between 60~K and 50~K.}
\label{fig8}
\end{figure}
to understand the nature of the LTO-LTT transition in this system. 
Axe {\it et al}.\cite{Axe1989} proposed a free energy near the structural phase transition 
from the LTO to the LTT (LTLO) phase with the expression of 
\begin{equation}
F(Q,\theta)=f(Q)+\alpha\cos4\theta+\beta[\cos8\theta-4\cos4\theta]
\label{eq2}
\end{equation}
where $\alpha\sim T_{\rm d2}-T$ near $T_{\rm d2}$. 
$Q$ and $\theta$ represent the amplitude of displacement and the rotation angle from 
$a$-axis for CuO$_{6}$ octahedra, which is defined 
in the inset of Fig.~\ref{fig1}. In this model, $\theta=0^{\circ}$ and $45^{\circ}$ 
correspond to the LTO and the LTT phase, respectively, and, in the case of 
$0^{\circ}<\theta<45^{\circ}$, the LTLO phase appears. 
Since the order parameter $Q$, which corresponds to 
the tilt amplitude of CuO$_{6}$ octahedra, is almost temperature independent in the vicinity 
of $T_{\rm d2}$, the free energy $F$ can be regarded as a function of only $\theta$. 
When $\beta>0$, Eq.~\ref{eq2} has 
one local minimum at $0^{\circ}<\theta<45^{\circ}$ below $T_{\rm d2}$ and 
the second-order phase transition from the LTO to LTLO phase occurs. On the other 
hand, for $\beta<0$, two local minimums at $\theta=0^{\circ}$ and $45^{\circ}$ 
appear in Eq.~\ref{eq2} and the first order LTO-LTT transition occurs, 
suggesting that the two phases can be coexist near $T_{\rm d2}$. 
Our present results and the previous study\cite{Keimer1993} of 
La$_{1.65}$Nd$_{0.35}$CuO$_{4}$ show the validity of this simple model. 
\section{Conclusion}\label{con}
The present study has confirmed the soft phonon behavior of 
La$_{1.875}$Ba$_{0.125}$CuO$_{4}$ in the wide temperature range 
including the HTT--LTO and the LTO--LTT structural phase transitions, 
which can be totally understood by soft-mode transitions. 
The dispersion relation of phonons along $a$-axis suddenly changes below $T_{\rm d2}$ 
while that along $c$-axis is independent of temperature, 
which might be characteristic in the LTT structure. 
The coexistence of the LTO and the LTT phase observed in both the phonon spectra and 
the Bragg peak profiles can be understood as the consequence of the first order 
structural phase transition in the frame work of Landau theory. 
The relevance between the change of the phonon velocity along $a$-axis in the LTT phase 
and the formation of charge stripe order should be clarified in the near future. 
\begin{acknowledgments}
The authors thank J.~M.~Tranquada, M.~H\"{u}cker and S.~Wakimoto 
for invaluable discussions. 
The neutron scattering experiments at JAERI was performed under the PACS 
No.~4474 and 4501. 
This work was supported in part by a Grant-In-Aid for Young 
Scientists B (15740194), 
Scientific research B (14340105), 
Scientific Research on Priority Areas (12046239), and 
Creative Scientific Research (13NP0201) from the Japanese Ministry of Education, 
Science, Sports and Culture, and by the US-Japan cooperative research program on 
Neutron scattering. Work at Brookhaven National Laboratory is supported by Division of 
Material Science, U.~S.~Department of Energy. 
\end{acknowledgments}


\begin{references}

\bibitem{Moodenbaugh1988}A.~R.~Moodenbaugh, Youwen~Xu, M.~Suenaga, T.~J.~Follcerts, 
and R.~N.~Shelton, 
Phys.~Rev.~B {\bf 38}, 4596 (1988).

\bibitem{Axe1989}J.~D.~Axe, A.~H.~Moudden, D.~Hohlwein, D.~E.~Cox, K.~M.~Mohanty, 
A.~R.~Moodenbaugh, and Youwen~Xu, 
Phys.~Rev.~Lett. {\bf 62}, 2751, (1989).

\bibitem{Crawford1991}M.~K.~Crawford, R.~L.~Harlow, E.~M.~McCarron, 
W.~E.~Farneth, J.~D.~Axe, H.~Chou, and Q.~Huang, 
Phys.~Rev.~B {\bf 44}, 7749 (1991).

\bibitem{Birgeneau1987}R.~J.~Birgeneau, C.~Y.~Chen, D.~R.~Gabbe, H.~P.~Jenssen, 
M.~A.~Kastner, C.~J.~Peters, P.~J.~Picone, Tineke Thio, T.~R.~Thurston, 
H.~L.~Tuller, J.~D.~Axe, P.~B\"{o}ni, and G.~Shirane, 
Phys.~Rev.~Lett. {\bf 59}, 1329, (1987).

\bibitem{Boni1988}P.~B\"{o}ni, J.~D.~Axe, G.~Shirane, R.~J.~Birgeneau, D.~R.~Gabbe, 
H.~P.~Jenssen, M.~A.~Kastner, C.~J.~Peters, P.~J.~Picone, and T.~R.~Thurston,
Phys.~Rev.~B {\bf 38}, 185 (1988).

\bibitem{Thurston1989}T.~R.~Thurston, R.~J.~Birgeneau, D.~R.~Gabbe, H.~P.~Jenssen, 
M.~A.~Kastner, P.~J.~Picone, N.~W.~Preyer, J.~D.~Axe, P.~B\"{o}ni, G.~Shirane, M.~Sato, 
K.~Fukuda, and S.~Shamoto, 
Phys.~Rev.~B {\bf 39}, 4327 (1989).

\bibitem{Keimer1993}B.~Keimer, R.~J.~Birgeneau, A.~Cassanho, Y.~Endoh, M.~Greven, 
M.~A.~Kastner, and G.~Shirane, 
Z.~Phys.~B {\bf 91}, 373 (1993). 

\bibitem{Kimura2004}$Q_{1}$ and $Q_{2}$ 
corresponds to displacements along $a$- and $b$-axis in the LTO phase, 
which arises from the tilting of CuO$_{6}$ octahedra. 
$Q$ can be defined as the vectorial sum of 
$Q_{1}$ and $Q_{2}$ with phase angle $\theta$. In this definitions, 
$\theta$ can be regarded as the angle between $a$-axis of the LTO phase and 
the tilt axis of CuO$_{6}$ octahedra, and $Q$ corresponds to the rotation amplitude 
around the tilt axis.

\bibitem{Braden1993}M.~Braden, O.~Hoffels, W.~Schnelle, B.~B\"{u}chner, 
G.~Heger, B.~Hennion, I.~Tanaka, and H.~Kojima, 
Phys.~Rev.~B {\bf 47}, 12288 (1993).

\bibitem{Lee1996}C.~H.~Lee, K.~Yamada, M.~Arai, S.~Wakimoto, S.~Hosoya, and Y.~Endoh, 
Physica~C {\bf 257}, 264 (1996).

\bibitem{Kimura2000}H.~Kimura, K.~Hirota, C.~H.~Lee, K.~Yamada, and G.~Shirane, 
J.~Phys.~Soc.~Jpn. {\bf 69}, 851 (2000).

\bibitem{Tranquada1995}J.~M.~Tranquada, B.~J.~Sternlieb, J.~D.~Axe, Y.~Nakamura, and S.~Uchida, 
Nature {\bf 375}, 561 (1995).

\bibitem{Tranquada1996}J.~M.~Tranquada, J.~D.~Axe, N.~Ichikawa, Y.~Nakamura, S.~Uchida, 
and B. Nachumi, 
Phys.~Rev.~B {\bf 54}, 7489 (1996).

\bibitem{Fujita2004}M.~Fujita, H.~Goka, K.~Yamada, J.~M.~Tranquada, and L.~P.~Regnault, 
Phys.~Rev.~B {\it in press} (cond-mat/0403396).

\bibitem{Billinge1993}S.~J.~L.~Billinge, G.~H.~Kwei, A.~C.~Lawson, and J.~C.~Thompson, 
Phys.~Rev.~Lett. {\bf 71}, 1903, (1993).

\bibitem{Ting1990}W.~Ting, K.~Fossheim, and T.~L\ae greid, 
Solid~State~Comm., {\bf 75}, 727 (1990).

\end{references}

\end{document}